\documentclass{aa}

\usepackage{graphicx}
\usepackage{epsfig}

\renewcommand{\d}{{\rm d}}
\newcommand{\inta}{\int_{-i\infty}^{+i\infty}}
\newcommand{\beq}{\begin{equation}}
\newcommand{\eeq}{\end{equation}}
\newcommand{\beqa}{\begin{eqnarray}}
\newcommand{\eeqa}{\end{eqnarray}}
\newcommand{\bea}{\begin{array}}
\newcommand{\ea}{\end{array}}
\newcommand{\cGN}{{\cal G}_{\rm N}}
\newcommand{\rhob}{\overline{\rho}}
\newcommand{\lag}{\langle}
\newcommand{\rag}{\rangle}
\newcommand{\Om}{\Omega_{\rm m}}
\newcommand{\Ol}{\Omega_{\Lambda}}

\newcommand{\dR}{\delta_{R}}
\newcommand{\br}{{\bf r}}
\newcommand{\bk}{{\bf k}}
\newcommand{\dL}{\delta_L}
\newcommand{\cF}{{\cal F}}
\newcommand{\cG}{{\cal G}}

\newcommand{\psib}{\overline{\psi}}
\newcommand{\DL}{\Delta_L}
\newcommand{\xiL}{\xi_L}
\newcommand{\PL}{P_L}
\newcommand{\cP}{{\cal P}}
\newcommand{\Det}{{\rm Det}}
\newcommand{\ti}{t_{\rm i}}
\newcommand{\ai}{a_{\rm i}}
\newcommand{\Ri}{R_{\rm i}}

\newcommand{\dLi}{\delta_{L,\rm i}}
\newcommand{\vi}{v_{\rm i}}
\newcommand{\Hi}{H_{\rm i}}
\newcommand{\thetai}{\theta_{\rm i}}
\newcommand{\etai}{\eta_{\rm i}}
\newcommand{\cPZ}{{\cal P}^Z}
\newcommand{\cFZ}{{\cal F}^Z}
\newcommand{\cGZ}{{\cal G}^Z}
\newcommand{\phiZ}{\varphi^Z}
\newcommand{\SZ}{S^Z}
\newcommand{\zi}{z_{\rm i}}
\newcommand{\Dp}{D_+}

\begin{document}

 

\title{Transients from Zel'dovich initial conditions}   
\author{P. Valageas}  
\institute{Service de Physique Th\'eorique, CEN Saclay, 91191 Gif-sur-Yvette, France} 
\date{Received / Accepted }

\abstract{
We investigate the error implied by the use of the Zel'dovich approximation to set up the initial conditions at a finite redshift $\zi$ in numerical simulations. Using a steepest-descent method developed in a previous work (\cite{paper2}) we derive the probability distribution $\cP(\dR)$ of the density contrast in the quasi-linear regime. This also provides its dependence on the redshift $\zi$ at which the simulation is started. Thus, we find that the discrepancy with the exact pdf (defined by the limit $\zi \rightarrow \infty$) is negligible after the scale factor has grown by a factor $a/\ai \ga 5$, for scales which were initially within the linear regime with $\sigma_{\rm i} \la 0.1$. This shows that the use of the Zel'dovich approximation to implement the initial conditions is sufficient for practical purposes since these are not very severe constraints. 
}

\maketitle

\section{Introduction}

In usual cosmological scenarios, large-scale structures in the universe form through the growth of small initial density fluctuations by gravitational instability (e.g., \cite{Peebles1}). Besides, in most cases of cosmological interest the amplitude of these perturbations increases at small scales, as in the CDM model (\cite{Peebles3}). This leads to a hierarchical scenario of structure formation where smaller scales become non-linear first. They build small virialized objects which later become part of increasingly large structures. Thus, this halos give rise to galaxies or clusters of galaxies (depending on cooling processes). Unfortunately, this non-linear regime is very difficult to handle analytically so that N-body simulations are a key tool to understand the formation of large-scale structures. They are even more important when one tries to follow the evolution of baryons which involves many processes (star formation, cooling, radiative ionization,..). Therefore, it is important to get a good estimate of the accuracy of such numerical simulations.

In this respect, an obvious source of error is the generation of initial conditions. Indeed, numerical simulations are initialized at a finite redshift $\zi$ while they should be started at time $\ti=0$ when the relevant scales are exactly ``linear''. In practice, one uses the Zel'dovich approximation (\cite{Zel1}) to set up the initial conditions at $\zi$. This correctly matches the exact density and velocity fields at linear order but the higher-order terms are not exact (as compared with the fields defined by the same linear growing mode initialized at $t=0$). This leads to a small error which can be made negligible by starting the simulations at a sufficiently large redshift $\zi$, when the scales of interest are far within the linear regime (e.g., \cite{Jus1}, \cite{Bau1}). 

These transients induced by the Zel'dovich approximation were investigated in \cite{Scoc1} for the first few order moments ($q \leq 8$) of the density and velocity fields using a perturbative approach. However, this approach involves lengthy calculations which worsen at higher orders and it is not obvious how to get an estimate of the error for the probability distribution function (pdf) $\cP(\dR)$ of the density contrast from these moments. Therefore, in this article we show how one can apply to this problem a steepest-descent method developed in a previous work (\cite{paper2}) which provides a rigorous derivation of $\cP(\dR)$ in the quasi-linear limit. This approach is quite general since it applies to Gaussian initial conditions (\cite{paper2}) as well as to non-Gaussian initial conditions (e.g., \cite{paper3}) or to the tails of $\cP(\dR)$ in both linear and non-linear regimes (\cite{paper4}). Also, it is actually quite intuitive. Here we describe how it can be used to derive the dependence on the initial redshift $\zi$ of the error induced on the pdf $\cP(\dR)$. It also yields the moments of the density field at any order.

This article is organized as follows. First, in Sect.\ref{Steepest-descent method} we recall the path-integral formulation which allows us to derive the pdf $\cP(\dR)$ in terms of initial conditions. Then, in Sect.\ref{Zel'dovich initial conditions} we apply this approach to the gravitational dynamics defined by Zel'dovich initial conditions at a finite redshift $\zi$. Finally, in Sect.\ref{Numerical results} we present our numerical results.

\section{Steepest-descent method}
\label{Steepest-descent method}

We first recall in this section the basis of the steepest-descent method built in \cite{paper2}. This also allows us to introduce our notation. We are interested in the statistics of the density contrast $\dR$ within a spherical cell of physical radius $R$, volume $V$:
\beq
\dR = \int_V \frac{\d^3r}{V} \; \delta(\br) .
\label{dR}
\eeq
Note that in this article all length-scales will be physical lengths (and not comoving coordinates). Here $\delta(\br)$ is the non-linear density contrast at the physical coordinate $\br$, at the time of interest. It is fully determined by the linear growing mode $\dL(\br)$ since we assume that the decaying mode has had time to vanish. Moreover, for Gaussian initial conditions the statistics of the linear density field $\dL(\br)$ are fully defined by the kernel $\DL(\br_1,\br_2)$:
\beq
\DL(\br_1,\br_2) \equiv \lag \dL(\br_1) \dL(\br_2) \rag .
\label{Dl}
\eeq
Note that $\DL(\br_1,\br_2)=\xiL(|\br_2-\br_1|)$ is simply the two-point correlation function of the linear density field, which is related to the linear power-spectrum $\PL(k)$ by:
\beq
\xiL(r) = \int \d\bk \; e^{i \bk.\br} \; \PL(k) .
\label{PL}
\eeq
Besides, the rms linear density fluctuation $\sigma(R)$ in a spherical cell of radius $R$, volume $V$, is:
\beq
\sigma^2(R) \equiv \lag \delta_{L,R}^2 \rag = \int_V \frac{\d\br_1}{V} \frac{\d\br_2}{V} \; \DL(\br_1,\br_2) ,
\label{sig1}
\eeq
where $\delta_{L,R}$ is the linear density contrast within the spherical cell of radius $R$. 

Next, in order to derive the moments and the pdf $\cP(\dR)$ of the density contrast it is convenient to introduce the generating function $\psib(y)$ defined by the relation:
\beq
\psib(y) \equiv \lag e^{-y \dR / \sigma^2(R) } \rag \equiv \int_{-1}^{\infty} \d\dR \; e^{-y \dR / \sigma^2(R) } \; \cP(\dR) .
\label{psib1}
\eeq
The average $\lag .. \rag$ in eq.(\ref{psib1}) denotes the mean over the Gaussian initial conditions, which are defined by the linear growing mode $\dL(\br)$. The last equality defines the pdf $\cP(\dR)$ of the density contrast. Thus, $\psib(y)$ is also the ``rescaled'' Laplace transform of $\cP(\dR)$.It also yields the moments of the density contrast $\dR$ through the expansion:
\beq
\psib(y) = \sum_{q=0}^{\infty} \frac{(- y)^q}{q!} \; \frac{\lag \dR^q \rag}{\sigma^{2q}} .
\label{psib2}
\eeq
It is convenient to introduce also the generating function $\varphi(y)$ of the cumulants defined by:
\beq
\psib(y) \equiv e^{-\varphi(y)/\sigma^2} ,
\label{phi1}
\eeq
which exhibits the expansion:
\beq
\varphi(y) = - \sum_{q=2}^{\infty} \frac{S_q}{q!} (- y)^q \;\; \mbox{with} \;\;S_q \equiv \frac{\lag \dR^q \rag_c}{\sigma^{2(q-1)}} ,
\label{phi2}
\eeq
since we have $\lag \dR \rag_c=0$. 

Then, the Gaussian average which appears in eq.(\ref{psib1}) can be expressed as the path-integral:
\beq
\psib(y) = \left( \Det\DL^{-1} \right)^{1/2} \int [\d\dL(\br)] \; e^{- S[\dL] /\sigma^2(R)}
\label{psib3}
\eeq
where we introduced the action $S[\dL]$:
\beq
S[\dL] \equiv y \; \dR[\dL] + \frac{\sigma^2(R)}{2} \; \dL . \DL^{-1} . \dL
\label{S1}
\eeq
Here $\dR[\dL]$ is the non-linear density contrast $\dR$ within the spherical cell $V$ obtained for a linear density field $\dL(\br)$. The action $S[\dL]$ is independent of the normalization of the power-spectrum $P_L(k)$ since $\DL \propto \sigma^2$. Then, it is clear that the path-integral in eq.(\ref{psib3}) is dominated by the minimum of the action $S$ in the limit $\sigma \rightarrow 0$ for a fixed $y$. Indeed, the contributions from other states $\dL$ are exponentially damped relative to this point. Moreover, the steepest-descent approximation becomes exact in this limit. This allows us to compute the generating functions $\varphi(y)$ and $\psib(y)$ in this limit. Finally, we obtain the pdf $\cP(\dR)$ from eq.(\ref{psib1}), through the inverse Laplace transform:
\beq
\cP(\dR) = \inta \frac{\d y}{2\pi i \sigma^2(R)} \; e^{[y \dR - \varphi(y)]/\sigma^2(R)} .
\label{P2}
\eeq

Next, in order to apply the steepest-descent method we need to derive the saddle-point of the action $S[\dL]$. As shown in \cite{paper2} a great simplification occurs thanks to the spherical symmetry of the problem. This symmetry arises because the initial conditions (i.e. the linear growing mode $\dL$) are statistically homogeneous and isotropic and we consider the density contrast $\dR$ within spherical cells. Then, one obtains a spherically symmetric saddle-point. Besides, for spherically symmetric states we can write:
\beq
\left\{ \begin{array}{l}
{\displaystyle \dR = \cF \left[ \delta_{L,R_L} \right] } \\ \\
{\displaystyle R_L^3 = (1+\dR) R^3 }
\end{array} \right.
\label{F1}
\eeq
where the function $\cF[\delta_{L,R_L}]$ is given by the usual spherical collapse solution of the equations of motion, see \cite{Peebles1} and \cite{paper2}. The second equation in (\ref{F1}) merely expresses the conservation of mass. Then, as shown in \cite{paper2} from eq.(\ref{F1}), for a power-law linear power-spectrum:
\beq
\PL(k) \propto k^n \;\;\; \mbox{with} \;\;\; -3<n<1 ,
\label{n1}
\eeq
the generating function $\varphi(y)$ given by the steepest-descent method is defined by the implicit system:
\beq
\left\{ \begin{array}{l}
{\displaystyle \tau = - y \; \cG'(\tau) } \\ \\
{\displaystyle \varphi(y) = y \; \cG(\tau) + \frac{\tau^2}{2} }
\end{array} \right.
\label{tau1}
\eeq
Here we introduced the function $\cG(\tau)$ which is related to the function $\cF[\delta_{L,R_L}]$ defined in eq.(\ref{F1}) by:
\beq
\cG(\tau) = \cF\left[ - \tau \; (1+\cG[\tau])^{-(n+3)/6} \right] .
\label{G1}
\eeq
This yields the pdf $\cP(\dR)$ through eq.(\ref{P2}) and eq.(\ref{tau1}). We refer the reader to \cite{paper2} for a detailed discussion of the steepest-descent method within this framework.

\section{Zel'dovich initial conditions}
\label{Zel'dovich initial conditions}

The results described in the previous section were derived for Gaussian initial conditions at the time $\ti=0$. Indeed, eq.(\ref{F1}) implies that $\dR/\delta_{L,R_L} \rightarrow 1$ for $t \rightarrow 0$. However, in numerical simulations the initial conditions are set up at a non-zero time $\ti>0$. This leads to a small error for the properties of the density field. This discrepancy vanishes in the limit $\ti \rightarrow 0$. Here we are interested in the quantitative estimate of this error for the moments and the pdf $\cP(\dR)$ of the density contrast. This can easily be done through the steepest-descent method recalled in Sect.\ref{Steepest-descent method}. Indeed, all the steps involved in the derivation of eq.(\ref{tau1}) and eq.(\ref{G1}) remain valid because the initial conditions are again statistically homogeneous and isotropic and they are still defined by the Gaussian linear growing mode $\dL(\br)$. However, the function $\cF[\delta_{L,R_L}]$ which describes the spherical dynamics is no longer given by the usual spherical collapse equations (i.e. eq.(39) and eq.(40) in \cite{paper2}). Therefore, in order to derive the pdf $\cP(\dR)$ we simply need to obtain the new function $\cF[\delta_{L,R_L}]$.

\subsection{Characteristic function $\cF[\delta_{L,R_L}]$}
\label{Characteristic function}

In the following we consider a critical-density universe where the analytic expression of the spherical dynamics is quite simple. However, our method obviously applies to any cosmological parameters. One simply needs to use the relevant expressions for the (time-dependent) function $\cF[\delta_{L,R_L}]$. Let us first consider the case of a spherical overdensity $\delta_{L,R_L}>0$. Thus, we study the evolution of a spherical region of constant mass $M$ and physical linear radius $R_L(t) \propto a(t)$, where we note $a(t)$ the scale-factor. We can take:
\beq
R_L(t) = a(t) \equiv t^{2/3} .
\label{RL1}
\eeq
The use of the Zel'dovich approximation to set up the initial conditions at $\ti>0$ does not break the spherical symmetry so that the equation of motion is given by the spherical dynamics:
\beq
\ddot{R} = - \frac{\cGN M}{R^2} .
\label{mot1}
\eeq
Here we introduced Newton's constant $\cGN$ and the actual physical radius $R(t)$ of the spherical overdensity. For a positive overdensity the general solution of eq.(\ref{mot1}) reads:
\beq
\left\{ \begin{array}{l}
{\displaystyle R = A (1-\cos \theta) } \\ \\
{\displaystyle t = C + B (\theta-\sin\theta) }
\end{array} \right. \;\;\; \mbox{with} \;\;\; A^3 = \cGN M B^2 .
\label{Rt1}
\eeq
Note that the offset $C$ in eq.(\ref{Rt1}) vanishes in the limit $\ti \rightarrow 0$ where we must recover the usual spherical dynamics. However, it must be taken into account in order to write the general solution to eq.(\ref{mot1}). In particular, the error introduced by the Zel'dovich approximation at $\ti>0$ yields a small non-zero value for this constant of integration $C$. 

Next, in order to obtain the function $\cF[\delta_{L,R_L}]$ we need to derive the parameters $A,B$ and $C$. First, we note that for a critical-density universe we have:
\beq
\rhob = \frac{1}{6\pi\cGN t^2}, \;\; \mbox{hence} \;\; M=\frac{2}{9\cGN} \;\; \mbox{and} \;\; A^3=\frac{2}{9} B^2 ,
\label{AB1}
\eeq
where we used eq.(\ref{RL1}). Thus, we can express the parameter $A$ in eq.(\ref{Rt1}) in terms of $B$. Next, we must specify the initial conditions at time $\ti$. We consider an overdensity with a linear density contrast $\dL$ at the time $t$ which we are interested in. At time $\ti$ the linear density contrast was $\dLi=(\ai/a) \dL$ (where we note $\ai=a(\ti)$ and $a=a(t)$ the scale-factors at times $\ti$ and $t$). The Zel'dovich approximation means that we apply to the particles the displacement given by linear theory. The exact radius $R$ obeys $R=R_L (1+\dR)^{-1/3}$ therefore the Zel'dovich approximation yields:
\beq
\Ri = \ai \left( 1 - \frac{1}{3} \dLi \right)
\label{Ri1}
\eeq
where we used the normalization (\ref{RL1}). From eq.(\ref{Ri1}) we see that the physical velocity $v(t) = \d R/\d t$ at time $\ti$ is:
\beq
\vi = \Hi \ai \left( 1 - \frac{2}{3} \dLi \right)
\label{vi1}
\eeq
within the Zel'dovich approximation, where we used $\dL(t) \propto a(t)$ and we introduced the Hubble constant $\Hi=H(\ti)$ at time $\ti$. Since $a(t) = t^{2/3}$ we also have: $H(t) = \dot{a}/a =2/(3t)$. The initial conditions (\ref{Ri1}) and (\ref{vi1}) at time $\ti$ define the constants of integration $B$ and $C$ in eq.(\ref{Rt1}) through:
\beq
\left\{ \begin{array}{l}
{\displaystyle \Ri = \left( \frac{2}{9} B^2 \right)^{1/3} (1-\cos \thetai) } \\ \\
{\displaystyle \ti = C + B (\thetai-\sin\thetai) } \\ \\
{\displaystyle \vi = \left( \frac{2}{9 B} \right)^{1/3} \frac{\sin \thetai}{1-\cos\thetai} }
\end{array} \right. 
\label{Rti1}
\eeq
This system also determines the parameter $\thetai$ associated with the time $\ti$. In particular, from eq.(\ref{Rti1}) we see that the combination $\Ri \vi^2$ only depends on the angle $\thetai$. Using eq.(\ref{Ri1}) and eq.(\ref{vi1}) we obtain:
\beq
\Ri \vi^2 = \frac{2}{9} (1+\cos\thetai) = \frac{4}{9} \left( 1 - \frac{1}{3} \frac{\ai}{a} \dL \right) \left( 1 - \frac{2}{3} \frac{\ai}{a} \dL \right)^2
\label{Rivi21}
\eeq
where we used $\dLi=(\ai/a) \dL$. This yields $\thetai$ as a function of $\dL$. Substituting into the first line of eq.(\ref{Rti1}), using eq.(\ref{Ri1}), we get:
\beqa
B & = & \left( \frac{2}{9} \right)^{-1/2} \left( \frac{10}{3} \right)^{-3/2} \left( \frac{\dL}{a} \right)^{-3/2} \left( 1 - \frac{1}{3} \frac{\ai}{a} \dL \right)^{3/2} \nonumber \\ & & \times \; \left( 1 - \frac{8}{15} \frac{\ai}{a} \dL + \frac{4}{45} \left(\frac{\ai}{a} \dL\right)^2 \right)^{-3/2} .
\label{B1}
\eeqa
Next, the angle $\theta$ which corresponds to the time $t$ of interest is given by:
\beq
t-\ti = B \left[ (\theta-\sin\theta) - (\thetai-\sin\thetai) \right]
\label{theta1}
\eeq
while the density contrast $\dR$ obeys:
\beq
1+\dR = \left( \frac{R_L}{R} \right)^3 = a^3 \left( \frac{2}{9} B^2 \right)^{-1} (1-\cos\theta)^{-3}.
\label{dR1}
\eeq
Putting together eq.(\ref{Rivi21}), eq.(\ref{theta1}) and eq.(\ref{dR1}), using eq.(\ref{B1}), we finally obtain, for $\dL \geq 0$:
\beq
\left\{ \begin{array}{l}
{\displaystyle \cos \thetai = 1 - \frac{10}{3} \frac{\ai}{a} \dL + \left(\frac{4}{3} \frac{\ai}{a} \dL\right)^2 - \left(\frac{2}{3} \frac{\ai}{a} \dL\right)^3 } \\ \\
{\displaystyle \theta-\sin\theta = \thetai-\sin\thetai + \left[ 1 - \left( \frac{\ai}{a} \right)^{3/2} \right] \left( \frac{2}{9} \right)^{1/2} } \\
{\displaystyle \hspace{1.7cm} \times \left( \frac{10}{3} \right)^{3/2} \dL^{3/2} \left( 1 - \frac{1}{3} \frac{\ai}{a} \dL \right)^{-3/2} } \\
{\displaystyle \hspace{1.7cm} \times \left( 1 - \frac{8}{15} \frac{\ai}{a} \dL + \frac{4}{45} \left(\frac{\ai}{a} \dL\right)^2 \right)^{3/2} } \\ \\
{\displaystyle 1+\dR = \left[ \frac{10}{3} \frac{ \dL \left( 1 - \frac{8}{15} \frac{\ai}{a} \dL + \frac{4}{45} \left(\frac{\ai}{a} \dL\right)^2 \right) }{ \left( 1 - \frac{1}{3} \frac{\ai}{a} \dL \right) (1-\cos\theta)} \right]^3 }
\end{array} \right.
\label{dRdL1}
\eeq
This system of equations fully defines the function $\cF[\delta_{L,R_L}]$ at time $t$ for positive overdensities since we have $\dR=\cF(\dL)$. Thus, the first line gives $\thetai$, the second one yields $\theta$ and the last one provides $\dR$. Note that for $\dL \rightarrow 0$ we have $\thetai \rightarrow 0$ and $\theta \rightarrow 0$. Moreover, since we consider the quasi-linear regime we can restrict ourselves to $\dL <3/2$ and the angle $\thetai$ obeys $0 \leq \thetai \leq \pi$ while we have $0 \leq \theta \leq 2\pi$. Of course, we can proceed in the same fashion for underdensities which yields, for $\dL \leq 0$:
\beq
\left\{ \begin{array}{l}
{\displaystyle \cosh \etai = 1 - \frac{10}{3} \frac{\ai}{a} \dL + \left(\frac{4}{3} \frac{\ai}{a} \dL\right)^2 - \left(\frac{2}{3} \frac{\ai}{a} \dL\right)^3 } \\ \\
{\displaystyle \sinh\eta-\eta = \sinh\etai-\etai + \left[ 1 - \left( \frac{\ai}{a} \right)^{3/2} \right] \left( \frac{2}{9} \right)^{1/2} } \\
{\displaystyle \hspace{1.7cm} \times \left( \frac{10}{3} \right)^{3/2} \left(-\dL\right)^{3/2} \left( 1 - \frac{1}{3} \frac{\ai}{a} \dL \right)^{-3/2} } \\
{\displaystyle \hspace{1.7cm} \times \left( 1 - \frac{8}{15} \frac{\ai}{a} \dL + \frac{4}{45} \left(\frac{\ai}{a} \dL\right)^2 \right)^{3/2} } \\ \\
{\displaystyle 1+\dR = \left[ - \frac{10}{3} \frac{ \dL \left( 1 - \frac{8}{15} \frac{\ai}{a} \dL + \frac{4}{45} \left(\frac{\ai}{a} \dL\right)^2 \right) }{ \left( 1 - \frac{1}{3} \frac{\ai}{a} \dL \right) (\cosh\eta-1)} \right]^3 }
\end{array} \right.
\label{dRdL2}
\eeq
Here we also have $\etai \geq 0$ and $\eta \geq 0$. Finally, as explained above, the determination of the function $\cF[\delta_{L,R_L}]$ also gives the pdf $\cP(\dR)$ through eq.(\ref{G1}), eq.(\ref{tau1}) and eq.(\ref{P2}).

\subsection{Zel'dovich pdf $\cPZ(\dR)$}
\label{Zel'dovich pdf}

At this point, it is interesting to consider the pdf $\cPZ(\dR)$ at the initial time $\ti$. This is in fact the pdf defined by the Zel'dovich approximation. It is directly obtained from eq.(\ref{Ri1}) and the first equality in eq.(\ref{dR1}) which gives (for any positive or negative $\dL$):
\beq
\cFZ[\delta_{L,R_L}] = \left( 1 - \frac{1}{3} \delta_{L,R_L} \right)^{-3} - 1 
\label{FZ1}
\eeq
where the superscript ``Z'' refers to the Zel'dovich approximation. Next, we can obtain the coefficients $\SZ_q$ defined in eq.(\ref{phi2}) by expanding the functions $\cFZ(\dL)$, $\cGZ(\tau)$ and $\phiZ(y)$ around zero. This yields:
\beq
\cGZ(\tau) = -\tau + \frac{1-n}{6} \tau^2 + .. , \;\; \phiZ(y) = - \frac{y^2}{2} + \frac{1-n}{6} y^3 + ..
\label{GZ1}
\eeq
which gives for the skewness:
\beq
\SZ_3 = 4-(n+3) .
\label{S3Z}
\eeq
Of course, we recover the results obtained in \cite{Kof1} who derived the parameters $S_q$ in the quasi-linear limit defined by the Zel'dovich approximation using perturbative means. Indeed, as shown in \cite{paper2} the steepest-descent method recalled in Sect.\ref{Steepest-descent method} gives the same results for these coefficients $S_q$ as the perturbative technique presented in Bernardeau (1992,1994). We refer the reader to \cite{paper2} (Sect.4.1) for a detailed comparison of the steepest-descent method with the perturbative approach.

Obviously, since the function $\cFZ(\dL)$ defined by the Zel'dovich approximation is different from the exact spherical function $\cF(\dL)$, the pdf $\cP(\dR)$ and the coefficients $S_q$ obtained at time $\ti$ differ from the value defined by the exact gravitational dynamics (where the initial conditions would be set at $t=0$). For instance, the skewness factor $\SZ_3$ obtained in eq.(\ref{S3Z}) is different from the exact value $S_3 = 34/7-(n+3)$ (e.g., \cite{Ber2}). However, as the initial conditions are set up at earlier times $\ti \ll t$ the error induced by the non-zero value of $\ti$ decreases and in the limit $\ti \rightarrow 0$ the skewness factor measured on linear scales at time $t$ tends towards the exact value $S_3 = 34/7-(n+3)$. The goal of the present article is precisely to study how fast this relaxation proceeds. In other words, we wish to see at which redshift $\zi$ we must set up the initial conditions so that the error due to the use of the Zel'dovich approximation at $\ti$ can be safely neglected. This can be obtained from the time-evolution of the function $\cF(\dL)$ derived in Sect.\ref{Characteristic function}.

\subsection{Lowest order moments}
\label{Lowest order moments}

The eq.(\ref{dRdL1}) and eq.(\ref{dRdL2}) provide a parametric definition of the function $\cF(\dL)$ at any time $t \geq \ti$. Besides, in the limit $\dL \rightarrow 0$ we have $\thetai \rightarrow 0$, $\theta \rightarrow 0$ (or $\etai \rightarrow 0$, $\eta \rightarrow 0$) and $\dR \rightarrow 0$. Therefore, by expanding the relations in eq.(\ref{dRdL1}) and eq.(\ref{dRdL2}) around zero we obtain the Taylor expansion of $\varphi(y)$ around $y=0$. This yields the coefficients $S_q$ through eq.(\ref{phi2}). These expansions are straightforward with a symbolic mathematical package (e.g., Mathematica), up to the required order. For instance, we give below the expansions up to the second-order which is needed for the skewness. We get:
\beq
\cF(\dL) = \dL + \left[ \frac{17}{21} - \frac{1}{5} \frac{\ai}{a} + \frac{2}{35} \left( \frac{\ai}{a} \right)^{7/2} \right] \dL^2 + .. 
\label{Fs1}
\eeq
and:
\beq
\cG(\tau) = -\tau + \left[ \frac{13}{42} - \frac{n}{6} - \frac{1}{5} \frac{\ai}{a} + \frac{2}{35} \left( \frac{\ai}{a} \right)^{7/2} \right] \tau^2 + .. 
\label{Gs1}
\eeq
which gives:
\beq
\varphi(y) = - \frac{y^2}{2} + \left[ \frac{13}{42} - \frac{n}{6} - \frac{1}{5} \frac{\ai}{a} + \frac{2}{35} \left( \frac{\ai}{a} \right)^{7/2} \right] y^3 + .. 
\label{phis1}
\eeq
and:
\beq
S_3 = \frac{34}{7} - (n+3) - \frac{6}{5} \frac{\ai}{a} + \frac{12}{35} \left( \frac{\ai}{a} \right)^{7/2} .
\label{S3s}
\eeq
We can check that our result (\ref{S3s}) agrees with the value obtained by \cite{Scoc1} from a perturbative approach. The advantage of our formulation is that it provides at once the coefficients $S_q$ up to any order. Besides, it also gives the pdf $\cP(\dR)$ through eq.(\ref{P2}). Of course, we can see that for $\ai=a$ we recover the predictions of the Zel'dovich approximation discussed in Sect.\ref{Zel'dovich pdf} while for $\ai \rightarrow 0$ we recover the exact values of the gravitational dynamics derived in \cite{paper2}.

\section{Numerical results}
\label{Numerical results}

Finally, we present in this section the numerical results obtained from the analysis described in Sect.\ref{Zel'dovich initial conditions}. We compare our results with the numerical simulations analysed in \cite{Ber2}. They use an adaptative P$^3$M code with $2.1\times 10^6$ particles. The error bars are estimated by dividing the simulation box in eight equal subsamples and by making eight different measurements.

\subsection{Skewness}
\label{Skewness}

\begin{figure}[htb]
\centerline{\epsfxsize=8cm \epsfysize=5.8cm \epsfbox{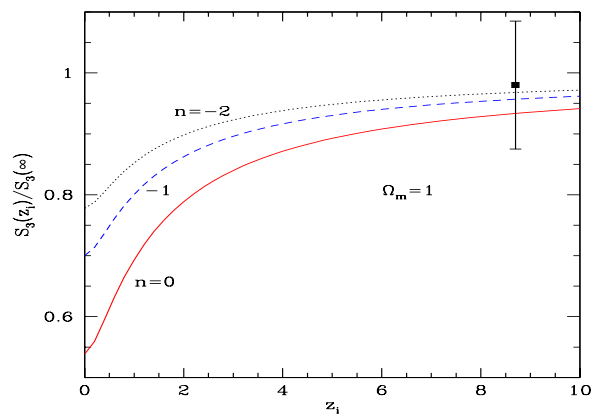}}
\caption{The dependence on $\zi$ of the ratio $S_3(\zi)/S_3(\infty)$ of the skewness factor $S_3$ measured at $z=0$ with respect to the exact value (obtained when $\zi=\infty$). We consider three linear power-spectra (i.e. $n=-2,-1$ and $n=0$, from top to bottom) in a critical-density universe. The coefficient $S_3$ is obtained from eq.(\ref{S3s}). The data point shows the measure of the skewness in a numerical simulation (\cite{Ber2}) with $\zi=8.7$ for $n=-1$.} 
\label{figS3}
\end{figure}

First, for completeness we display in Fig.\ref{figS3} the dependence of the skewness factor $S_3$ measured today (i.e. at $z=0$) with the redshift $\zi$ at which the initial conditions are set up. We consider three linear power-spectra: $n=-2,-1$ and $n=0$. The curves show the ratio $S_3(\zi)/S_3(\infty)$, where $S_3(\infty)$ is also the exact value of the skewness factor on quasi-linear scales (i.e. when the initial conditions are set up at $t=0$). Of course, we can check that the error over $S_3$ decreases as $\zi \rightarrow \infty$. Besides, we can note that the discrepancy with the exact value grows for larger $n$. The data point shows the measure of $S_3$ in the numerical simulation for $n=-1$ after the scale-factor has grown by a factor $a/\ai=9.7$ (i.e. $\zi=8.7$). We can see that the error due to the Zel'dovich approximation at the initial time is already negligible, as compared with the accuracy limitations due to finite size effects. 

As seen in \cite{Scoc1} the error also increases for larger order $q$ of the parameters $S_q$. Note that these results only hold on large scales with $\sigma \la 0.8$, which belong to the quasi-linear regime. These results were already obtained in \cite{Scoc1} from a perturbative approach hence we shall not comment them further here.

\subsection{Probability distribution $\cP(\dR)$}
\label{Probability distribution}

\begin{figure}[htb]
\centerline{\epsfxsize=8cm \epsfysize=5.8cm \epsfbox{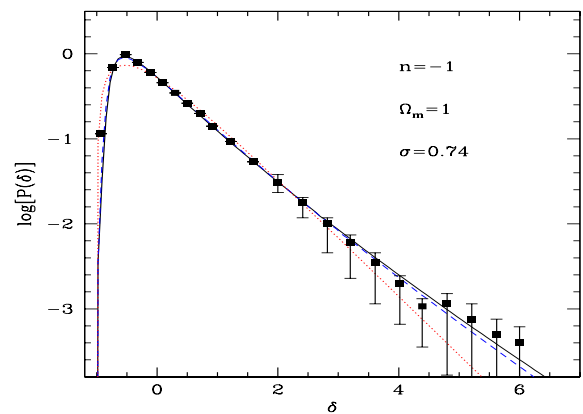}}
\caption{The pdf $\cP(\dR)$ for $n=-1$, $\Om=1$ and $\sigma=0.74$, at redshift $z=0$. The solid line shows the ``exact'' pdf (i.e. when the initial conditions are set up at $\zi=\infty$). The dashed line corresponds to $\zi=5$ and the dotted line to $\zi=0$ (i.e. to the Zel'dovich approximation). The data points (obtained from numerical simulations) are taken from \cite{Ber2} with $\zi=15$.} 
\label{figPn1}
\end{figure}

We can note that the lowest order moments do not fully describe the pdf $\cP(\dR)$ so that it is not easy to get an accurate estimate of the deviation of $\cP(\dR)$ from its asymptotic value from the first few coefficients $S_q$. Fortunately, as detailed in Sect.\ref{Zel'dovich initial conditions} we also derived the pdf $\cP(\dR)$ itself. Thus, we present in Fig.\ref{figPn1} the dependence on $\zi$ of the pdf $\cP(\dR)$ measured today at $z=0$, in a critical-density universe with $n=-1$. The case $\zi=0$ actually gives the reconstructed pdf implied by the Zel'dovich approximation. We can clearly see the difference with the actual pdf which comes from the exact gravitational dynamics with initial conditions set up at $\ti=0$ (obtained within the quasi-linear approximation). However, we note that at $\zi=5$ the pdf is already very close to this asymptotic value and at $\zi=10$ it would be nearly indistinguishable in the figure. Therefore, we can conclude that for practical purposes the scales which obey $\sigma_{\rm i} \la 0.1$ at the time $\ti$ when the initial conditions are set in are well described by the numerical simulation after the scale-factor has grown by a factor $a/\ai \ga 5$. In particular, it is clear from Fig.\ref{figPn1} that, provided these constraints are satisfied, the error bars of the numerical simulations are largely dominated by other problems (e.g., finite size effects, finite resolution). Note that the measures obtained from the simulations correspond to $\zi=15$ where the error due to the set-up of the initial conditions is indeed negligible.

For simplicity we have only considered the case of a critical-density universe so far. Nevertheless, it is clear that our approach can be extended to any cosmology. One simply needs to use the relevant equations of motion which describe the spherical dynamics of the saddle-point (e.g., eq.(\ref{Rt1})). This merely changes the characteristic function $\cF[\delta_{L,R_L}]$. However, this is not necessary for practical purposes. Indeed, since observations show that the present universe is close to the critical density ($\Om \ga 0.2$) at high redshift we have $\Om(z) \simeq 1$ to a very good accuracy. Moreover, in order to study large-scale structures in the present universe numerical simulations must start sufficiently early at $z>20$, especially if one is interested in the behaviour of baryons which involves early processes like reionization at $z\ga 6$. Therefore, during the initial stages of realistic simulations we have indeed $\Om(z) \simeq 1$.

Nevertheless, in case one is only interested in the present universe at $z \sim 0$ it is important to take into account the effect of the cosmological parameters on the time-evolution of the transients due to the Zel'dovich approximation. To do so, one must use the relevant characteristic function $\cF[\delta_{L,R_L}]$ which is no longer given by eq.(\ref{dRdL1}) and eq.(\ref{dRdL2}). For instance, if $\Ol >0$ an additional term appears in the equation of motion (\ref{mot1}) which we must integrate. However, as noticed in \cite{Nuss1} the evolution of density perturbations is almost independent of the cosmological parameters if we use the linear growth factor $\Dp(t)$ as a time variable. This means that the results of Fig.\ref{figPn1} remain valid except that the redshift $\zi=5$ shown by the dotted line actually corresponds to the redshift $\zi'$ given by:
\beq
\frac{\Dp(0)}{\Dp(\zi')} = 1+\zi ,
\label{zip1}
\eeq
where we used the fact that for the critical-density universe we have $\Dp \propto a \propto (1+z)^{-1}$. For instance, for a low-density flat universe with $\Om=0.3$ and $\Ol=0.7$ we get:
\beq
\zi=5 \;\; \rightarrow \;\; \zi'=6.7 \;\; \mbox{for} \;\; \Om=0.3 , \Ol=0.7
\label{zip2}
\eeq
Thus, as expected we find that a larger number of expansion factors are required in the case of low-density universes as structure formation is frozen at late time by the more rapid expansion of the universe.

\section{Conclusion}

Thus, in this article we have shown how to derive the pdf $\cP(\dR)$ of the density contrast implied by the use of the Zel'dovich approximation to set up the initial conditions at a finite redshift $\zi$. This allows us to obtain a rigorous quantitative estimate of the error due to this effect for the pdf $\cP(\dR)$ and the moments of the density contrast. Our results apply to large scales which are still in the quasi-linear regime. This is quite sufficient for practical purposes since it is clear that it makes no sense to use the Zel'dovich approximation at time $\ti$ for scales which are already non-linear.

Then, we found that the error for the pdf $\cP(\dR)$ associated with the Zel'dovich approximation at $\zi$ is negligible after the scale factor has grown by a factor $a/\ai \ga 5$, for scales which were initially within the linear regime with $\sigma_{\rm i} \la 0.1$. In low-density universes one needs a larger expansion factor ratio $a/\ai$. In fact, to a good approximation the relevant quantity is the growth factor ratio $\Dp(t)/\Dp(\ti)$, which should be larger than $5$. Thus, in order to use numerical simulations it is important to check that these constraints are indeed satisfied over the scales one is interested in. However, these conditions are not very severe, hence it appears that this effect is not the main source of error. Therefore, it is sufficient to use the Zel'dovich approximation to set up initial conditions and one does not need to use higher-order approximations.

\begin{acknowledgements}

We would like to thank F. Bernardeau for discussions which motivated this work.

\end{acknowledgements}

\end{document}